\documentclass[pre,twocolumn,showpacs,superscriptaddress,amsmath]{revtex4}
\usepackage{amsfonts,amssymb,bm,pifont}
\usepackage{times,mathptm}
\usepackage[dvips]{graphicx}

  \usepackage{color}

\begin{document}

%



\title{Numerical Study of a Microscopic Artificial Swimmer}

\author{Erik Gauger}
\affiliation{%
Fachbereich Physik, Universit\"at Konstanz,
D-78457 Konstanz, Germany
}
\author{Holger Stark}
\affiliation{%
Fachbereich Physik, Universit\"at Konstanz, D-78457 Konstanz, Germany
}
\affiliation{%
Max-Planck-Institut f\"ur Dynamik und Selbstorganisation,
Bunsenstr. 10, D-37073 G\"ottingen, Germany}

\date{\today}

\begin{abstract}
We present a detailed numerical study of a microscopic artificial swimmer
realized recently by Dreyfus {\em et al.} in experiments
[R. Dreyfus {\em et al.\/}, Nature\ \textbf{437}, 862 (2005)].
It consists of an elastic filament composed of superparamagnetic particles 
that are linked together by DNA strands. Attached to a load particle, the
resulting swimmer is actuated by an oscillating external magnetic field
so that it performs a non-reciprocal motion in order to move forward. 
We model the superparamagnetic filament by a bead-spring configuration
that resists bending like a rigid rod and whose beads experience friction
with the surrounding fluid and hydrodynamic interactions with each other.
We show that, aside from finite-size effects, its dynamics is governed 
by the dimensionless sperm number, the magnitude of the magnetic field,
and the angular amplitude of the field's oscillating direction. Then 
we study the mean velocity and the  efficiency of the swimmer
as a function of these parameters and the size of the load particle.
In particular, we clarify that the real velocity of the swimmer is 
influenced by two main factors, namely the shape of the beating filament 
(determined by the sperm number and the magnetic-field strength)
and the oscillation frequency. Furthermore, the load size influences
the performance of the swimmer and has to be chosen as a compromise between 
the largest swimming velocity and the best efficiency.  Finally, we 
demonstrate that the direction of the swimming velocity changes in a 
symmetry-breaking transition when the angular amplitude of the field's 
oscillating direction is increased, in agreement with experiments.
\end{abstract}


\maketitle

\section{Introduction} \label{intro}
Nature was very inventive to design mechanisms that microorganisms
such as bacteria and many eukaryotic cells use to propel themselves in 
a highly viscous environment, i.~e., at very low Reynolds numbers\ 
\cite{Bray2001}. Since 
they cannot rely on drifting by inertia, as we do when we swim in water, 
they immediately come to a halt when they stop with their beating motion.
In this article we study in detail a microscopic artificial swimmer\
\cite{Dreyfus05}
that was constructed recently on the basis of a superparamagnetic 
elastic filament that mimics the so-called flagellum employed by 
many eukaryotic cells\ \cite{Bray2001}.

In 1977 Purcell pointed out in his famous article 
``Life at low Reynolds number'' that microorganisms have to perform
a non-reciprocal periodic motion to be able to move forward\ 
\cite{Purcell77} (see also\ \cite{Berg93}). Non-reciprocal means
that the time-reversed motion is not the same as the original one
(for examples see\
\cite{Shapere87,Stone96,Becker03,Avron04,Najafi04,Dreyfus05a}).
The reason lies in the Stokes equations\ \cite{Dhont1996}
governing the fluid flow around the microorganisms for negligible inertia: 
they allow for a time-inverted flow pattern when all the external forces are
inverted. 

Bacteria employ a marvelous rotary motor to crank a relatively
stiff helical filament\ \cite{Berg73,Berg04}. Spermatozoa as one 
example for eukaryotic cells move forward by creating bending 
waves that move along their elastic flagella from the head 
to the tail\ \cite{Taylor51,spermatozoa,Linck01}. 
While these waves are generated by
a collective motion of internal molecular motors\cite{Camalet99,Camalet00}, 
Dreyfus {\em et al.\/} use an external magnetic field to induce the 
beating of a superparamagnetic filament attached to a red-blood cell\ 
\cite{Dreyfus05}. The filaments are made from 
superparamagnetic colloidal particles of micron size. A static external 
magnetic field induces dipoles in the colloids so that they form a
chain. In the gaps between the charged colloids chemical linkers such as
double-stranded DNA are attached to the particles and an elastic filament
resisting bending and stretching is formed\ \cite{Goubault03,Cohen05,Koenig05}
(for similar systems see\ \cite{Biswal}). 
As demonstrated by the impressive experiments of Dreyfus {\em et al.\/}, 
an oscillating external magnetic field now induces a non-reciprocal beating 
motion of the superparamagnetic filament that is able to move the 
attached red-blood cell forward.

So far, the modelling of the dynamics of the superparamagnetic filament
followed the elastohydrodynamics of an elastic rod\ 
\cite{Wiggins98,Camalet00}
supplemented by a continuum version for the interaction of the
magnetic-field induced dipoles\ \cite{Dreyfus05,Roper05} or a simpler
description for the interaction with the magnetic field\ \cite{Cebers}. 
The authors are able to describe the dynamics of the 
filament\ \cite{Roper05,Cebers} and the velocity curve of the 
artificial swimmer\ \cite{Dreyfus05}.
Here we present a different description of the artificial swimmer 
similar to the one used by Lagomarsino and Lowe for driven 
microfilaments\ \cite{Lowe03,Lagomarsino04}. 
We take into account the discrete nature of the superparamagnetic filament 
by modelling it as a sequence of beads, and, in contrast to the work of 
Roper {\em et al.}\ \cite{Roper05}, we consider dipolar and hydrodynamic 
interactions between all the beads.
We present a thorough investigation of the swimming velocity and
efficiency as a function of the relevant parameters, study the influence 
of the size of the load attached to the filament, and demonstrate that the
swimmer can take different directions via a symmetry-breaking transition
depending on how the actuating magnetic field oscillates.

In Sec.\ \ref{sec.model} we present details for the modeling of the
dynamics of the superparamagnetic filament and show that aside from
finite-size effects it is governed by a few relevant parameters.
Section\ \ref{sec.swim} summarizes and discusses the results from our 
numerical study and Sec.\ \ref{sec.conc} contains our concluding 
remarks.

\section{Modelling the Superparamagnetic Filament} \label{sec.model}

\begin{figure}
\includegraphics[width=0.95\columnwidth]{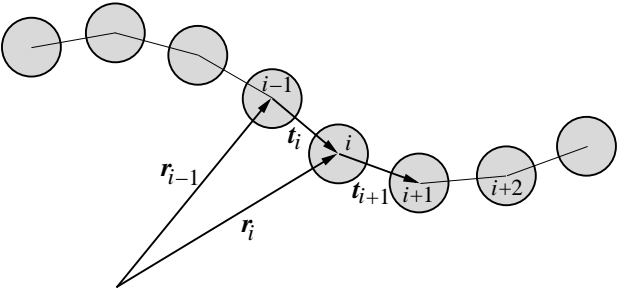}
\caption{Bead-spring model for the superparamagnetic filament consisting of
$N$ particles. Each bead at position $\bm{r}_{i}$ experiences forces due to
bending and stretching and due to interactions of the magnetic-field induced
dipoles located on the beads.}
\label{fig1}
\end{figure}

We model the superparamagnetic filament by a bead-spring configuration, 
as illustrated in Fig.\ \ref{fig1} that, in addition, resists bending 
like a worm-like chain\ \cite{wormlike}. Thus each bead in the filament 
experiences a force 
due to stretching, bending and dipolar interactions. While the chemical
linkers between the beads are responsible for the resistance to
bending and stretching, we completely ignore their contribution to
hydrodynamic friction. So the filament interacts with the fluid
surrounding via the hydrodynamic friction of the beads that
incorporates their hydrodynamic interactions. In the following
we first set up the discretized version of the bending and stretching
free energies\ \cite{twisting}
and also formulate the dipolar interaction energy
from which we then calculate the forces on a single bead.
In a second step, the equations of motion for the single beads are 
established with the help of their mobilities on the Rotne-Prager
level. Comments on numerical details and the scaling behavior conclude
this section.

\subsection{Energies and Forces} \label{subsec.energies}

A deviation of the distance $l_{i}=|\bm{t}_{i}|$ of the beads from their
equilibrium value $l_{0}$ obeys Hooke's law and the total stretching
free energy is
\begin{equation}
H^S = \frac{1}{2}  k  \sum_{i=2}^{N}  (l_i - l_0)^2 \, ,
\label{2.1}
\end{equation}
where $N$ denotess the total number of beads.
Typically, we consider relatively stiff springs so the variation
of $l_{i}$ is always smaller than $0.1l_{0}$. The stretching force acting on 
bead $i$ obeys
\begin{equation}
\bm{F}^S_i = - \bm{\nabla}_{\bm{r}_{i}} H^{S} = 
- k (l_i - l_0) \hat{\bm{t}}_i + k (l_{i+1} - l_0) \hat{\bm{t}}_{i+1} \, ,
\label{2.2}
\end{equation}
where $\bm{\nabla}_{\bm{r}_{i}}$ is the nabla operator with respect
to $\bm{r}_{i}$.

The linkers connecting the beads convey some bending rigidity to the
filament. The bending free energy of an elastic rod or a worm-like
chain is given by\ \cite{Landau91}
\begin{equation}
H^B = \frac{1}{2} A \int_0^L ds \left( \frac{d \hat{\bm{t}}}{d s} \right)^2 \,,
\label{2.3}
\end{equation}
where $L$ is the total length of the filament, $s$ the arclength along it,
and $\hat{\bm{t}}$ the unit tangent at location $s$. Refering the bending 
stiffness $A= k_{\text{B}}T l_{p}$ to the thermal energy $k_{\text{B}}T$, 
one obtains the persistence length $l_{p}$ that gives the length scale
on which the filament becomes flexible. Replacing $d \hat{\bm{t}}/(ds)$
by  $(\hat{\bm{t}}_{i+1} - \hat{\bm{t}}_{i}) / l_{0}$, we arrive at the
discretized version of the bending energy for the superparamagnetic
filament,
\begin{equation}
H^B = \frac{A}{l_0} \sum_{i=1}^{N} f_{i} (1 - \hat{\bm{t}}_{i+1} \cdot 
\hat{\bm{t}}_i) \, ,
\label{2.4}
\end{equation}
where $\hat{\bm{t}}_i = \bm{t}_i / l_{i}$. We have introduced the factor
\begin{equation}
f_i = \left\{ 
   \begin{array}{l}
       1 \enspace \text{for} \enspace 2 \leq i \leq N-1\\
       0 \enspace \text{for} \enspace i=1,N
   \end{array} \right.
\label{2.5}
\end{equation}
to let the index $i$ in Eq.\ (\ref{2.4}) still run from 1 to $ N$, i.~e.,
over all beads.
This facilitates the calculation of the bending force $\bm{F}^B_i =
- \bm{\nabla}_{\bm{r}_{i}} H^{B}$ acting on each bead $i$:
\begin{eqnarray}
\lefteqn{
\bm{F}^B_i = \frac{A}{l_0} \left\{ \frac{f_{i-1}}{l_i} \hat{\bm{t}}_{i-1} 
- \left[ \frac{f_{i-1}}{l_{i}} \hat{\bm{t}}_{i-1} \cdot \hat{\bm{t}}_{i} + 
\frac{f_{i}}{l_{i+1}} + \frac{f_{i}}{l_{i}} \hat{\bm{t}}_{i} \cdot 
\hat{\bm{t}}_{i+1} \right] \hat{\bm{t}}_i \right.} \nonumber \\
& & \left. + \left[ \frac{f_{i}}{l_{i+1}} \hat{\bm{t}}_{i} \cdot 
\hat{\bm{t}}_{i+1} + \frac{f_{i}}{l_{i}} + \frac{f_{i+1}}{l_{i+1}} 
\hat{\bm{t}}_{i+1} \cdot \hat{\bm{t}}_{i+2}\right] \hat{\bm{t}}_{i+1} 
- \frac{f_{i+1}}{l_{i+1}} \hat{\bm{t}}_{i+2} \right\} \quad \enspace
\label{2.6}
\end{eqnarray}
To arrive at $\bm{F}^B_i$ we have used the relations 
$\bm{\nabla}_{\bm{r}_i} \hat{\bm{t}}_i = 
\frac{1}{l_i}(\bm{1} - \hat{\bm{t}}_i \otimes \hat{\bm{t}}_i)$ and  
$\bm{\nabla}_{\bm{r}_i} \hat{\bm{t}}_{i+1} = -\frac{1}{l_{i+1}}
(\bm{1} - \hat{\bm{t}}_{i+1} \otimes \hat{\bm{t}}_{i+1})$ where $\bm{1}$
is the unit tensor and the symbol $\otimes$ means tensor product.

%
%
A single particle made from material with magnetic susceptibility $\chi$
and subjected to an external magnetic field $\bm{B}$ (more correctly
the magnetic induction) aquires a magnetic dipole moment
\begin{equation}
\bm{p} = \frac{4 \pi a^3}{3 \mu_0} \chi \bm{B} \,,
\label {2.7}
\end{equation}
where $\mu_{0} = 4\pi \cdot 10^{-7} \mathrm{N/A^{2}}$ is the permeability 
of free space and $a$ is the particle's radius. In the filament, the dipole 
moments $\bm{p}_{i}$ of the beads interact with each other with a total 
dipole-dipole interaction energy
\begin{equation}
H^{D} = \frac{\mu_0}{4\pi} \sum_{i,j=1}^{N}\hspace*{-0.8ex}^{\prime} \, 
 \frac{\bm{p}_i \cdot \bm{p}_j - 3 (\bm{p}_i \cdot \hat{\bm{r}}_{ij}) 
(\bm{p}_j \cdot \hat{\bm{r}}_{ij})}{r_{ij}^3} \, .
\label{2.8}
\end{equation}
where $r_{ij} = |\bm{r}_{j}-\bm{r}_{i}|$, $\hat{\bm{r}}_{ij} = 
(\bm{r}_{j}-\bm{r}_{i})/r_{ij}$ and $\sum^{\prime}$ means sum over all 
$i \ne j$. When all the beads experience
the same magnetic field $\bm{B} = B \hat{\bm{p}}$, the total interaction
energy becomes
\begin{equation}
H^{D} = \frac{4 \pi a^{6}}{9\mu_0}(\chi B)^{2} 
\sum_{i,j=1}^{N}\hspace*{-0.8ex}^{\prime} \,
 \frac{1-3(\hat{\bm{p}} \cdot \hat{\bm{r}}_{ij})}{r_{ij}^3} \, .
\label{2.9}
\end{equation}
Here we ignore that the local magnetic field differs from 
the external $\bm{B}$ since all the induced dipoles contribute to the local
field.
Taking into account the field from the nearest neighbors of a dipole, 
one can show that this renormalizes
the susceptibility $\chi$ and gives it a small anisotropy with
$\chi_{\|} - \chi_{\perp} \approx -0.25$ for a typical value of 
$\chi\approx 1$, where $\|$ and $\perp$ refer, respectively, to directions 
parallel and perpendicular to the local filament axis\ \cite{Roper05}.
In the following we are interested in the basic features of the system
and therefore work with the energy (\ref{2.9}) to calculate the
force $\bm{F}^D_i = - \bm{\nabla}_{\bm{r}_i} H^{D}$ that all magnetic
dipoles excert on bead $i$:
\begin{equation}
\bm{F}^D_i = \frac{4 \pi a^{6}}{3\mu_0}(\chi B)^{2}
\sum_{j=1}^{N}\hspace*{-0.5ex}^{\prime}
\frac{2(\hat{\bm{p}} \cdot \hat{\bm{r}}_{ij})\hat{\bm{p}} +
[1-5(\hat{\bm{p}} \cdot \hat{\bm{r}}_{ij})^{2}]\hat{\bm{r}}_{ij}}{r_{ij}^{4}}
\, .
\label{2.10}
\end{equation}
The filament is now driven by an external field whose direction 
$\hat{\bm{p}}(t)$ and magnitude $B(t)$ depend on time. For example, if 
the direction is oscillating very slowly, the filament will follow the
field like a rigid rod since the induced dipoles prefer to align themselves
along a common line. 
However, if the oscillations become faster, the hydrodynamic
friction of the beads with the surrounding fluid increases so that they
are not fast enough to form a straight line and the filament is
bending.

\subsection{Equations of Motion}
The filament is immersed in a viscous fluid such as water. On the length 
scale of microns and for times larger than the momentum relaxation time,
inertia can be neglected so that the velocities $\bm{v}_{i}$ of the beads
are proportional to the forces acting on them\ \cite{Dhont1996}. 
Hence the beads follow the equations of motion
\begin{equation}
\bm{v}_i = \sum_{j} \bm{\mu}_{ij} \bm{F}_{j} \enspace \mathrm{with} 
\enspace \bm{F}_{j} = \bm{F}_j^{S} + \bm{F}_j^{B} + \bm{F}_j^{D} \, ,
\label{2.11}
\end{equation}
where the forces $\bm{F}_j^{S}$, $\bm{F}_j^{B}$, and $\bm{F}_j^{D}$
are given in the last section. They all depend on the beads´locations
$\bm{r}_{i}$ and the dipolar forces also possess an explicite time
dependence. The important quantities in our treatment are the mobilities 
$\bm{\mu}_{ij}$. Roper {\em et al.\/} considered the filament as a 
continuous line whose friction with the surrounding fluid is governed 
locally by two anisotropic friction coefficients for respective motions 
parallel and perpendicular to the local direction of the filament\ 
\cite{Roper05}. In our treatment, the
anisotropic friction results from hydrodynamic interactions between
neighboring beads. Moreover, hydrodynamic interactions between more
distant beads are also taken into account. Since induced flow fields
are long ranged (they decay as $1/r$, where $r$ is the distance from a 
moving bead), this effect cannot be neglected. In general, 
hydrodynamic interactions constitute a complicated many-body problem\ 
\cite{Dhont1996}, however, their leading contribution is given by 
two-particle interactions. Moreover, 
if the particles are not too close to each other so that lubrication 
becomes important, the Rotne-Prager approximation can be employed\ 
\cite{Dhont1996,Rotne}. We use it in a version for spheres with different 
radii $a_{i}$ and $a_{j}$\ \cite{Jeffrey83}.
Whereas all spheres in the filament have the same radius 
$a_{i} = a$ ($1 \le i \le N$), we also attach a larger sphere
with radius $a_{0}$ to the filament to mimic its load. In the Rotne-Prager 
approximation, the self mobility is simply
\begin{equation}
\bm{\mu}_{ii} = \mu_{0} \bm{1} \enspace \mathrm{with} \enspace
\mu_{0} = 1/(6\pi\eta a_{i}) \, ,
\label{2.12}
\end{equation}
and the cross mobilities read
\begin{equation}
\bm{\mu}_{ij}  =  \frac{1}{6 \pi \eta r_{ij}} \left[ \frac{3}{4}
(\bm{1} + \hat{\bm{r}}_{ij} \otimes \hat{\bm{r}}_{ij})
+ \frac{1}{4}  \frac{a_i^2 + a_j^2}{r_{ij}^2} 
(\bm{1} - 3 \hat{\bm{r}}_{ij} \otimes \hat{\bm{r}}_{ij}) \right] \,.
\label{2.13}
\end{equation}

To obtain the particle paths $\bm{r}_{i}(t)$ and thus the dynamics of
the filament, we numerically integrate Eqs.\ (\ref{2.11}) using the Euler 
method\ \cite{Recipes92}. More accurate
schemes such as the Runge-Kutta method that would allow larger time
steps during the integration and thus speed up the simulations
are not useful since the maximum time step for the integration is
governed by the relaxation dynamics of local bending and stretching
deformations. At the time step chosen to avoid numerical instabilities,
the performance of the Euler method is comparable, e.~g., to the 
fourth-order Runge-Kutta scheme. In concrete, the relative differences
of the particle positions calculated with both methods are smaller
than $10^{-4}$.

\subsection{Reduced Equations of Motion}
Our modelling of the filament comprises several parameters. To identify
the essential parameters that govern the dynamics of the filament, we
rescale the dynamic equations appropriately so that only reduced variables 
appear. For example, all lengths will be referred to the equilibrium length of 
the filament that we approximate by $L \approx N l_{0}$.

To arrive at reduced equations of motion, we first rescale the energies
from section\ \ref{subsec.energies} with the help of characteristic
quantities:
\begin{equation}
H^{S} = \frac{1}{2} k \frac{L^{2}}{N} \, \tilde{H}^{S} \enspace, \enspace
H^{B} = \frac{A}{L} \, \tilde{H}^{B} \enspace, \enspace
H^{D} = \frac{4 \pi a^{6}}{9\mu_0}(\chi B)^{2} \frac{N^{4}}{L^{3}} \,
\tilde{H}^{D}
\label{2.14}
\end{equation}
where the reduced energies read:
\begin{eqnarray}
\tilde{H}^{S} & = &  
   N \sum_i \left(\frac{l_i - l_{0}}{L}\right)^2 
\label{2.15} \\
\tilde{H}^{B} & = &  
   N \sum_i (1 - \hat{\bm{t}}_i \cdot \hat{\bm{t}}_{i+1})
\label{2.16}  \\
\tilde{H}^{D} & = & \frac{1}{N^4} \sum_{i \ne j}  
\frac{1 - 3(\hat{\bm{r}}_{ij} \cdot \hat{\bm{p}})^2}{(r_{ij}/L)^{3}} \, .
\label{2.17}
\end{eqnarray}
The prefactor $kL^{2}/N$ of $\tilde{H}^{S}$ in Eqs.\ (\ref{2.14}) is 
essentially the stretching modulus of an elastic rod and $\tilde{H}^{S}$
averages the square of the strain variable over the whole rod.
The bending free energy of the filament with a uniform curvature $L/\sqrt{2}$
amounts to $A/L$ and the prefactor of $\tilde{H}^{D}$ gives the order
of magnitude of the energy of $N$ interacting dipoles. To compare
these characteristic energies to each other, we introduce the reduced
magnetic field $B_{s}$\ \cite{footnote} and the parameter $k_{s}$:
\begin{eqnarray}
B_{s} & = & \left(
\frac{\frac{4 \pi a^{6}}{9\mu_0}(\chi B)^{2} \frac{N^{4}}{L^{3}} }{A/L}
\right)^{1/2} 
= \,\, \frac{2\pi^{1/2}a^{3}\chi N}{3\mu_{0}^{1/2}l_{0} A^{1/2}}B 
\label{2.18} \\
k_{s} & = & \frac{kL^{2}/N}{A/L} \,\, = \,\, \frac{N^{2}l_{0}^{3}}{A} k \, .
\label{2.19}
\end{eqnarray}
In the following, we will characterize the strength of the magnetic field
by $B_{s}$, the parameter $k_{s}$ will basically be kept constant.

Whereas the scaling factors in Eqs.\ (\ref{2.14}) determine the magnitude
of the energies, the reduced energies of Eqs.\ (\ref{2.15})-(\ref{2.17})
distinguish between different types of stretching and bending deformations
and arrangements of dipoles along the filament. For sufficiently large
$L \approx N l_{0}$, they tend to constant values $\tilde{H}^{S}_{\infty}, 
\tilde{H}^{B}_{\infty}$, and $\tilde{H}^{D}_{\infty}$. 
We checked how finite size
effects from the ends of the filaments influence these constant values.
Bending the filament to a semicircle, the deviation from 
$\tilde{H}^{B}_{\infty}$
is around 1\% when $N = 20$, a typical number of beads in the 
superparamagnetic filament both in experiments and in our simulations.
On the other hand, for the same number of beads $\tilde{H}^{D}$ 
for dipoles aligned along a straight line deviates 
from $\tilde{H}^{D}_{\infty}$ by around 10\% due to the long range nature of 
the dipolar interaction.

To formulate the reduced equations of motion, we rescale length, time,
and velocity according to
\begin{equation}
\tilde{\bm{r}} = \bm{r}/L \enspace, \enspace
\tilde{t} = \omega t \enspace, \enspace
\frac{d \tilde{\bm{r}}}{d \tilde{t}} = \frac{1}{\omega L}
\frac{d \bm{r}}{d t} \, ,
\label{2.20}
\end{equation}
where $\omega$ is the frequency of the actuating magnetic field.
The reduced forces and mobilities follow from
\begin{equation}
\widetilde{\bm{F}}_{i} = \frac{\bm{F}_{i}}{A/L^{2}} \enspace, \enspace
\tilde{\bm{\mu}}_{ij} = 6\pi \eta a \frac{L}{l_{0}} \bm{\mu}_{ij} \, .
\label{2.21}
\end{equation}
Note that $6\pi \eta a L / l_{0}$ is the total friction coefficient of the
filament without taking into account hydrodynamic interactions.
Applying the rescaled quantities to Eqs.\ (\ref{2.11}) and using
Eqs.\ (\ref{2.14}), (\ref{2.18}), and (\ref{2.19}), one arrives
at the reduced equation of motion for bead $i$:
\begin{equation}
\frac{d \tilde{\bm{r}}_{i}}{d \tilde{t}} = S_{p}^{-4}
\sum_{j} \tilde{\bm{\mu}}_{ij} 
\widetilde{\bm{F}}_{j} 
\label{2.22}
\end{equation}
with
\begin{equation}
\widetilde{\bm{F}}_{j} = -\bm{\nabla}_{\tilde{\bm{r}}_{j}}
(\tilde{H}^{B} + k_{s} \tilde{H}^{S}/2 + B_{s}^{2} \tilde{H}^{D}) \, .
\end{equation}
The important parameter $S_{p}$ in Eqs.\ (\ref{2.22}), introduced first 
by Wiggins and Goldstein\ \cite{Wiggins98,Wiggins98a} and termed
{\em sperm number\/} by Lowe\ \cite{Lowe03}, is defined via
\begin{equation}
S_{p} = \left(\frac{6\pi\eta \frac{a}{l_{0}} \omega L^{4}}{A} \right)^{1/4}
= \frac{L}{l_{h}}
\enspace.
\label{2.23}
\end{equation}
It compares the frictional to the bending forces and 
completely determines the dynamics of an elastic filament in
a viscous environment in the so-called resistive-force theory
of slender bodies\ \cite{Brennen77}, 
where the friction with the surounding fluid
is described by two local friction coefficients per length, 
$\gamma_{\|}$ and $\gamma_{\perp}$, for respective motions
parallel and perpendicular to the local axis of the filament.
Our definition of $S_{p}$ agrees with the one of Wiggins and Goldstein\ 
\cite{Wiggins98,Wiggins98a} when we make the reasonable identification 
$\gamma_{\perp} = 6\pi\eta a / l_{0}$. An immediate interpretation
of $S_{p} = L/l_{h}$ is given with the help of the elastohydrodynamic 
penetration length $l_{h}$\ \cite{Wiggins98,Wiggins98a}. 
Consider a sufficiently long filament ($L \gg l_{h}$) whose one end 
undergoes an oscillation with frequency $\omega$. Then $l_{h}$ is the 
length on which the oscillation penetrates into the filament. 
On the other hand, if $L \ll l_{h}$, the filament oscillates as a whole 
like a rigid rod.

Other important quantities are the strength of the actuating magnetic field 
quantified by the parameter $B_{s}$ and its time protocol, which we will
concretize in section\ \ref{sec.swim}. The reduced stretching constant $k_{s}$ 
is always chosen such that variations in the length of the filament are
smaller than 10\%.

\begin{figure}
\includegraphics[width=0.95\columnwidth]{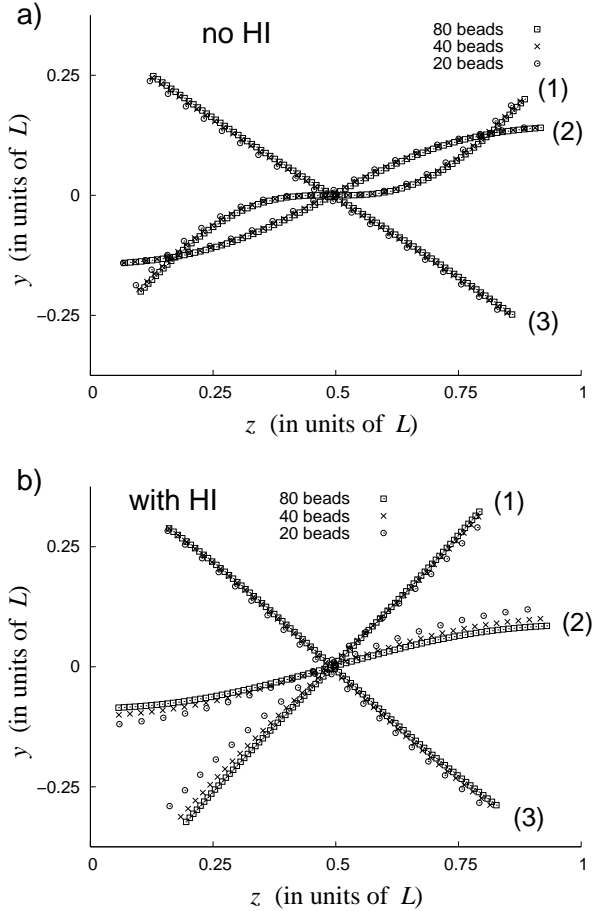}
\caption{Configurations of the superparamagnetic filament at the same moments
within one period of the oscillating magnetic field. The reduced parameters
 $S_{p}, B_{s}$, and $k_{s}$ are kept constant while the number of beads is
changed. a) without hydrodynamic interactions, b) with hydrodynamic 
interactions. The numbers at the snapshots indicate the pair of 
variables $\varphi,\omega t/(2\pi)$ for the time protocol of the 
oscillating magnetic field, as illustrated in Fig.\ \ref{fig3}: 
(1) $50^{\circ},0.25$; (2) $0^{\circ},0.5$; (3) $-35^{\circ},0.88$.}
\label{fig2}
\end{figure}

We already mentioned that the dynamics of a continuous elastic filament 
described within the resistive-force theory is completely determined by 
the sperm number $S_{p}$, which especially incorporates the influence 
from the filament length. In the superparamagnetic filament simulated by us 
through a discretized model, finite-size effects are present, as already 
discussed for the bending and dipolar energies. Hydrodynamic interactions 
are of longer range than dipolar interactions and we therefore discuss 
shortly how variations in the length of the filament or in the number of 
beads influence the dynamics for constant $S_{p}$, $B_{s}$, and $k_{s}$.
We keep the strength of the magnetic field constant but let its 
direction oscillate in the $yz$ plane around the $z$ axis with an 
angular amplitude of $50^{\circ}$. We solve the equations of 
motion\ (\ref{2.11}) for real parameters and change the length of the
filament for constant equilibrium-bead distant $l_{0}$ by choosing
the particle numbers $N=20$, 40, and 80. 
The reduced parameters are kept fixed at the values
$S_{p}=7.7$, $B_{s}=6.6$, and $k_{s}=4500$ by changing, respectively,
the frequency $\omega$ of the oscillating magnetic field, its strength $B$, 
and the real spring constant $k$. 
Figure\ \ref{fig2}a) shows snapshots of the different
filaments in the $yz$ plane at the same moments within the period of the 
oscillating field. Hydrodynamic interactions are completely switched off.
For $N=40$ and 80 the configuration of the filaments are basically
identical whereas for $N=20$ one realizes small differences due to
the finite-size effects from the dipolar interactions as discussed
above. In Fig.\ \ref{fig2}b) the configurations are shown at the
same moments within the period as in Fig.\ \ref{fig2}a) but now with
hydrodynamic interactions switched on. They clearly have a pronounced
effect on the oscillating shape of the filament. Since the beads move
collectively through the fluid, hydrodynamic interactions reduce, 
roughly speaking the local friction with the surrounding fluid.
The filament can more easily follow the oscillating magnetic field
direction and, as a result, it is less bent. The finite-since effect
is more pronounced compared to Fig.\ \ref{fig2}a) but does not
change the configurations significantly. So when in the following all our
simulations are done with $N=20$ to reduce computing time, we are sure
that the dynamics of the filament is mainly described by the
sperm number $S_{p}$, the reduced strength $B_{s}$ of the magnetic field, and
its time protocol.

\section{Study of the Artificial Swimmer} \label{sec.swim}

\begin{figure}
\includegraphics[width=0.95\columnwidth]{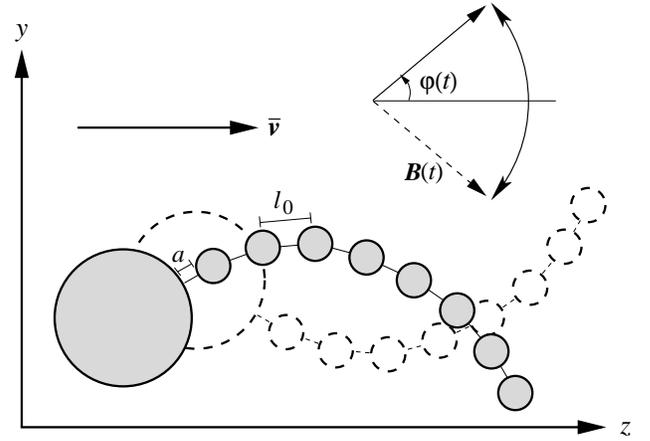}
\caption{For the artificial swimmer, a larger load particle is attached
to the filament. It is actuated by a magnetic field whose direction
is oscillating about the $z$ axis with 
$\varphi(t) = \varphi_{\mathrm{max}} \sin(\omega t)$. The two 
configurations of the swimmer are schematic drawings for positive and 
negative $\varphi(t)$.}
\label{fig3}
\end{figure}

The actuated superparamagnetic filament in Fig.\ \ref{fig2} does not 
move on average in one direction since it performs a reciprocal motion. This 
symmetry is broken when a load is attached to one end of the filament
as demonstrated by the wonderful experiments of Dreyfus {\em et al.\/} where
the load is a red-blood cell\ \cite{Dreyfus05}. In the following we study 
such an artificial swimmer in detail. As illustrated in  Fig.\ \ref{fig3},
we actuate the filament with a magnetic field $\bm{B}(t)$ whose
strength is constant but whose direction oscillates about the $z$ axis with 
an angle $\varphi(t) = \varphi_{\mathrm{max}} \sin(\omega t)$ (this
time protocol differs from the one used in Ref.\ \cite{Dreyfus05}).
As a result, the swimmer will move with an average velocity $\bar{\bm{v}}$ 
along the $z$ axis. However, in contrast to spermatozoa, where the
head is pushed forward by damped waves travelling from the head to the
tail\ \cite{Taylor51,spermatozoa}, 
the superparamagnetic filament drags the passive load behind 
itself by performing a sort of paddle motion with its free end as 
indicated in Fig.\ \ref{fig3}. 
Note that the direction can be reversed for certain parameters when the 
load particle also becomes superparamagnetic, but we will not study this
possibility further. In our approach, hydrodynamic interactions between the 
constituent beads are essential for the artificial swimmer to move forward.
In particular, they produce local friction coefficients $\gamma_{\|}$ and 
$\gamma_{\perp}$ for respective motions parallel and perpendicular to the 
filament's local axis that are not equal, $\gamma_{\|} \ne \gamma_{\perp}$.
This is a necessary prerequisit for swimming\ \cite{Lagomarsino04}. 
To implement hydrodynamic interactions, we use the Rotne-Prager approximation 
for the mobilities as introduced in Eqs.\ (\ref{2.12}) and\ (\ref{2.13}) 
in the form that also allows for hydrodynamic interactions between the 
filament's beads and the larger load particle. We simulate the artficial
swimmer by solving the equations of motion\ (\ref{2.11}) numerically for
two sets of parameters summarized in Appendix\ \ref{app.par}. The first
set is close to the realization of the artificial swimmer in 
Ref.\ \cite{Dreyfus05}. Since we already now that the filament itself
is governed by the dimensionless parameters $S_{p}$, $B_{s}$, and $k_{s}$, 
we present our results as functions of the sperm number $S_{p}$ and the
reduced field strength $B_{s}$. In addition, we also show that the 
angular amplitude $\varphi_{\mathrm{max}}$ of the magnetic-field 
oscillations and the size $a_{0}$ of the load particle influence the 
dynamics of the swimmer.

We discuss the performance of the swimmer by studying in detail its
average speed $\bar{\bm{v}}$ and its efficiency to move a load.
The speed $\bar{\bm{v}}$ is calculated by averaging the velocity 
$\bm{v}_{0}$ of the load particle over one actuation cycle:
\begin{equation}
\bar{\bm{v}} = \frac{1}{T} \int_{\tau}^{\tau+T} \bm{v}_{0}(t) dt \, ,
\label{2.24}
\end{equation}
where $T=2\pi/\omega$. In order to obtain a reliable $\bar{\bm{v}}$
from Eq.\ (\ref{2.24}), it is important that the swimmer has reached
a steady state at time $\tau$ so that the swimming does not depend any
more on initial conditions. For small frequencies ($S_{p} \approx 1$),
this state is reached rather quickly after one actuation cycle but
needs several cycles when $S_{p}$ is increased. Secondly, we define the 
efficiency of the swimmer to transport a load by comparing the energy
dissipated by the load, when moved uniformly with velocity $\bar{\bm{v}}$,
to the total energy dissipated by the swimmer:
\begin{equation}
\xi = \frac{6\pi \eta a_{0} \bar{v}^{2}}{\overline{\sum_{i=0}^{N}
\bm{F}_{i} \cdot \bm{v}_{i}}} \, ,
\end{equation}
where the bar in the denominator means average over one actuation cycle.
The efficiency $\xi$ indicates how much energy from the total energy used
to actuate the swimmer is employed to move the load particle forward with 
velocity $\bar{\bm{v}}$.

\begin{figure}
\includegraphics[width=0.95\columnwidth]{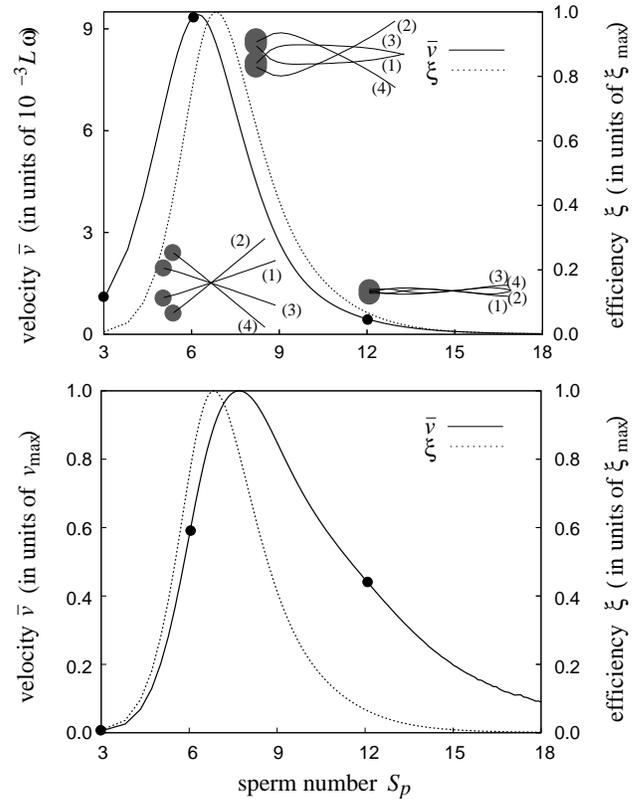}
\caption{Swimming velocity $\bar{v}$ and efficiency $\xi$ in units
of $\xi_{\mathrm{max}} = 1.58 \cdot 10^{-3}$ as a function of 
sperm number $S_{p}$ for reduced magnetic-field strength $B_{s} = 5.76$.
Upper graph: reduced velocity $\bar{v}/(L\omega)$, lower graph: 
absolut velocity $\bar{v}$ in units of 
$v_{\mathrm{max}} = 5.56 \cdot 10^{-5} \mathrm{m/s}$. The insets show several
snapshots of the filament's configuration for $S_{p} =3, 6$ and 12, 
respectively, indicated by the dots. The numbers at the snapshots 
indicate the pair of variables $\varphi,\omega t/(2\pi)$ for the time 
protocol of the oscillating magnetic field, as illustrated in 
Fig.\ \ref{fig3}: (1) $20^{\circ},0.07$; (2) $40^{\circ},0.17$; (3) 
$-20^{\circ},0.57$; (4) $-40^{\circ},0.67$. Parameter set I was used in 
simulations.}
\label{fig4}
\end{figure}

In Fig.\ \ref{fig4}, we plot the mean velocity $\bar{v} = |\bar{\bm{v}}|$ 
and the efficiency $\xi$ as a function of the sperm number $S_{p}$. 
For the simulations, the realistic parameter set I was used. The curve 
for $\bar{v}/(L\omega)$  in the upper graph of Fig.\ \ref{fig4} follows 
the same behavior as observed in Ref.\ \cite{Lowe03,Lagomarsino04},
where the elastic filament is driven by an oscillating torque acting on 
one of its ends. For small sperm numbers around $S_{p} = 3$, the reduced 
velocity is small since the superparamagnetic filament behaves nearly 
like a rigid rod, as illustrated by the inset on the lower left. 
It shows snapshots of the filament for $S_{p} = 3$. As already mentioned,
the oscillating motion of a rigid rod is reciprocal and therefore cannot 
produce a directed motion of the swimmer. Increasing the sperm number, e.~g., 
by increasing the frequency $\omega$, speeds up the artificial 
swimmer. At the maximum value of $\bar{v}/(L\omega)$ at around $S_{p} =6$, 
the increased friction with the surrounding fluid is able to bend the 
whole filament (see upper inset) which obviously promotes a high swimming 
velocity. Further increase in $S_{p}$ leads to a decrease in 
$\bar{v}/(L\omega)$; due to the strong friction with the fluid the
whole filament cannot follow the magnetic field and only a small wiggling 
of its free end remains. The efficiency as a function of $S_{p}$ shows 
a similar behavior as $\bar{v}/(L\omega)$ that is not completely
surprising: oscillating a rigid rod (small $S_{p}$) or fast wiggling of the 
filament (large $S_{p}$) dissipates energy but does not produce an effective 
motion. So one expects a maximum of $\xi$ close to the maximum of 
$\bar{v}/(L\omega)$ since $\xi$ is determined by $\bar{v}^{2}$.
Note that the small absolute efficiency ($\xi_{\mathrm{max}} = 
1.58 \cdot 10^{-3}$) is due to the fact that
a large amount of energy is dissipated by the motion of the filament.
The shape of the velocity curve changes when absolut velocities are plotted,
as demonstrated in the lower graph of Fig.\ \ref{fig4}. At $S_{p}=3$, the
absolute velocity is nearly zero (due to the small frequency) and the 
maximum is shifted to a larger value around $S_{p}=7.5$. Interestingly, the 
absolute velocities of the oscillating filaments at $S_{p}=6$ and $12$ do not 
differ so much, as first implied by the upper graph. This suggests that 
the absolute swimming velocity depends on two factors: (1) the shape of the 
oscillating filament, where bending the whole filament favors large 
velocities, and (2) the oscillation frequency. The latter point leads to the
very slow decrease of $\bar{v}$ as a function of $S_{p}$ in the lower graph. 
A comparison with the narrow maximum of $\xi$, however, shows that a lot 
of energy is dissipated at large $S_{p}$ in the small wiggling motion of the 
filament. So operating the artificial swimmer at around $S_{p} = 7$
between the two close maxima ensures highest swimming velocities with very 
efficient energy consumption.

\begin{figure}
\includegraphics[width=0.95\columnwidth]{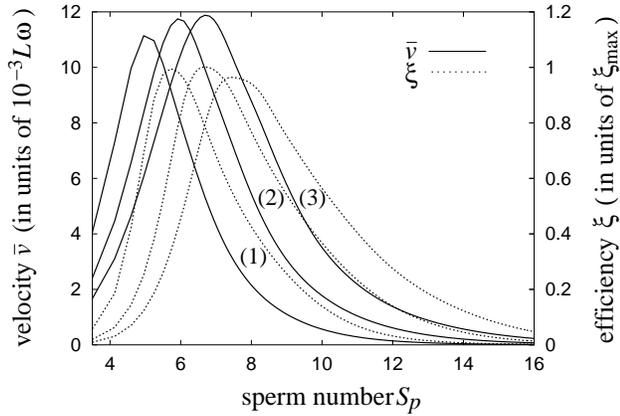}
\caption{Reduced swimming velocity $\bar{v}/(L\omega)$ and efficiency 
$\xi$ in units of $\xi_{\mathrm{max}} = 1.87 \cdot 10^{-3}$ as a function of 
sperm number $S_{p}$ for several reduced magnetic-field strengths: 
(1) $B_{s} = 4.2$, (2) $B_{s} = 6.07$, and (3) $B_{s} = 7.93$.
Parameter set II was used in simulations.}
\label{fig5}
\end{figure}

\begin{figure}
\includegraphics[width=0.95\columnwidth]{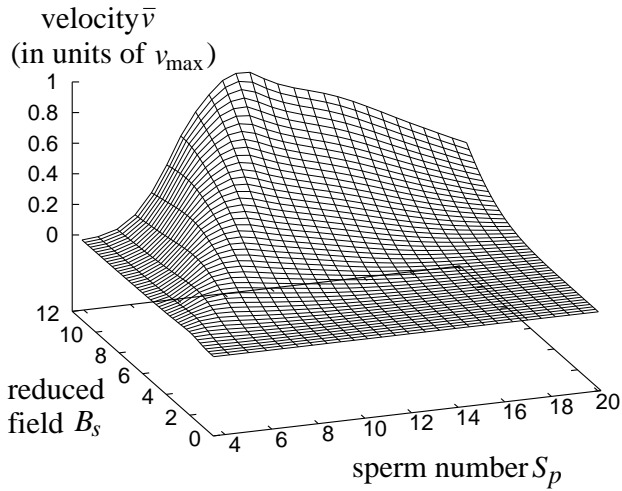}
\caption{Absolute swimming velocity $\bar{v}$ in units of the maximal value
$v_{\mathrm{max}}$ as a function of sperm number $S_{p}$ and
reduced magnetic-field strength $B_{s}$. Parameter set II was used
in simulations.}
\label{fig6}
\end{figure}

In Fig.\ \ref{fig5} we present reduced velocities and efficiencies 
as a function of $S_{p}$ for several reduced field strengths $B_{s}$,
which we simulated with parameter set II. We observe that with increasing 
$B_{s}$ the maxima of both the velocity and the efficiency curves move to 
larger sperm numbers. We understand this since larger magnetic fields mean 
stronger alignment of the dipoles and, therefore, larger resistance to 
bending. Both, maximium velocity and efficiency, only exhibit
a small dependence on $B_{s}$. After a slight increase of the maximum
velocity, it slightly decreases to a constant value (not shown in the graph).
Note that the reduced velocities are larger by a factor of 1.2 compared
to Fig.\ \ref{fig4} that we attribute to the larger angular amplitude
$\varphi_{\mathrm{max}} = 57^{\circ}$ compared to 
$\varphi_{\mathrm{max}} = 45^{\circ}$ used in 
Fig.\ \ref{fig4}. Finally, Fig.\ \ref{fig6} summarizes the dependence
of the absolut swimming velocity on sperm number and reduced field
calculated with parameter set II. For increasing field strength at constant
$S_{p}$, the graph shows a strong increase of the swimming velocity. At 
small $S_{p}$ a subsequent saturation at a constant value is visible. 
Note that the maximum velocity for constant $B_{s}$ is shifted to higher
sperm numbers when $B_{s}$ increases. This explains why the
the maximum reduced velocity $\bm{v}/(L\omega)$ in Fig.\ \ref{fig5} does not 
exhibit a strong dependence on $B_{s}$.

\begin{figure}
\includegraphics[width=0.95\columnwidth]{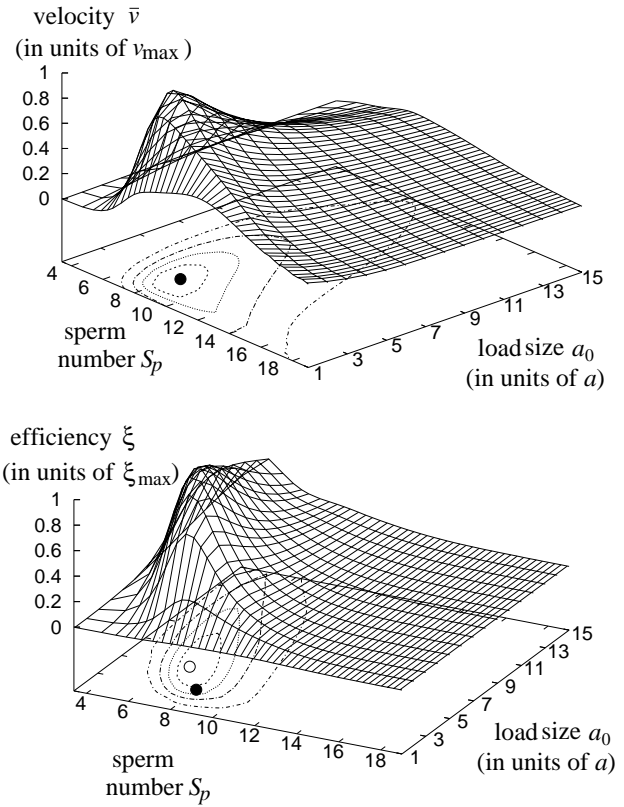}
\caption{Swimming velocity $\bar{v}$ 
  (in units of $v_{\mathrm{max}} = 7.31 \cdot 10^{-5} \mathrm{m/s}$)  
and efficiency $\xi$ (in units of $\xi_{\mathrm{max}} = 1,54\cdot 10^{-3}$)
as a function of sperm number $S_{p}$ and load size $a_{0}$ in units of $a$.
The filled and open circles indicate, respectively, the absolute
maxima of $\bar{v}$ and $\xi$. $B_{s} = 5.76$ and parameter set I was used 
in simulations.}
\label{fig7}
\end{figure}

The swimming velocity $\bar{v}$ clearly depends on the size or the 
radius $a_{0}$ of the load particle. In the upper graph of 
Fig.\ \ref{fig7}, we plot its absolute value as a function of sperm number 
$S_{p}$ and $a_{0}$ for a constant field strength $B_{s} = 5.76$ calculated
with parameter set I. There is a pronounced maximum at $S_{p} =8$ and 
$a_{0} \approx 3a$ (indicated by a filled circle), 
where $a$ is the radius of the 
superparamagnetic beads in the filament. The velocity $\bar{v}$ decreases
for large $a_{0}$ since the load becomes too heavy to be efficiently 
moved by the oscillating filament.
On the other hand, when $a_{0}$ approaches $a$, we also expect small 
swimming velocities since the asymmetry of the swimmer becomes small.
However, even at $a_{0}=a$ the swimming velocity is not zero and therefore
the swimmer still performs a non-reciprocal motion. The reason is that 
the load particle is not superparamagnetic and, therefore, a small
asymmetry remains. The plot for the reduced velocity $\bar{v}/(L\omega)$
looks similar to the lower graph of Fig.\ \ref{fig7}; the absolute maximum 
is not that pronounced and the maxium for constant $a_{0}$ moves to smaller 
sperm numbers when $a_{0}$ is increased. The lower graph shows how the 
efficiency $\xi$ behaves as a function of $S_{p}$ and $a_{0}$. The absolute 
maximum, indicated by an open circle, is at $S_{p}=6.6$ and $a_{0} \approx 5a$.
For comparison the location of the maximum of the swimming velocity is 
again shown by the filled circle. Since it is relatively sharp, one has to 
choose a compromise for the performance of the swimmer between the largest
swimming velocity and the best efficiency. The location of the maximum
velocity certainly depends on the magnetic-field strength. We expect 
that it is shifted to larger radii $a_{0}$ when $B_{s}$ increases. 
However, we have not studied this dependence in detail.

\begin{figure}
\includegraphics[width=0.9\columnwidth]{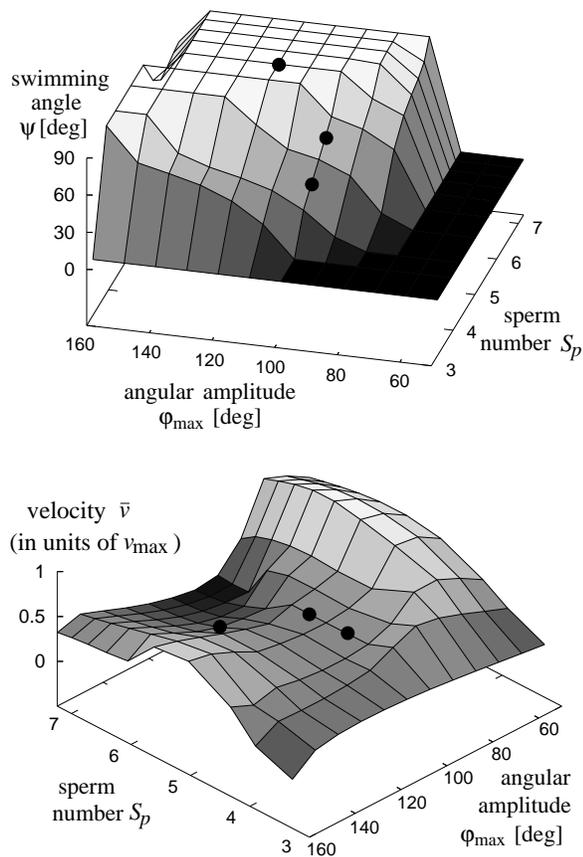}
\caption{Swimming velocity  $\bar{v}$ 
(in units of $v_{\mathrm{max}} = 5.57 \cdot 10^{-3} \mathrm{m/s}$)
and the swimming angle $\psi$ (measured relative to the $z$ axis)
as a function of sperm number $S_{p}$ and angular
amplitude $\varphi_{\mathrm{max}}$ of the oscillating field.
$B_{s} = 3.81$ and parameter set II was used in simulations.}
\label{fig8}
\end{figure}

\begin{figure}
\includegraphics[width=0.85\columnwidth]{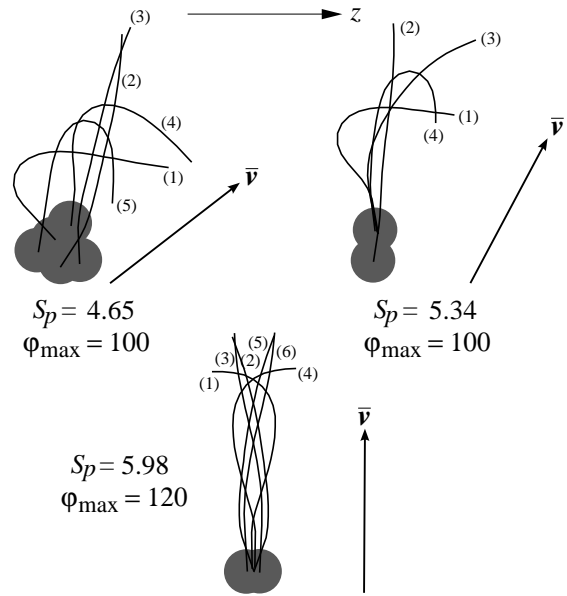}
\caption{Snapshots of the swimmer's configurations for different
parameters giving rise to different swimming directions. Their locations
in the graphs of Fig.\ \ref{fig8} are indicated by filled black circles.
The numbers at the snapshots indicate the pair of variables 
$\varphi,\omega t/(2\pi)$ for the time protocol of the oscillating magnetic 
field. $S_{p} =4.65$: (1) $0^{\circ},0.00$; (2) $95^{\circ},0.20$; (3) 
$58^{\circ},0.40$; (4) $-58^{\circ},0.60$; (5) $-95^{\circ},0.80$.
$S_{p} = 5.34$: (1) $0^{\circ},0.00$; (2) $100^{\circ},0.25$; (3) 
$0^{\circ},0.50$; (4) $-100^{\circ},0.75$.
$S_{p} = 5.98$: (1) $41^{\circ},0.06$; (2) $118^{\circ},0.22$; (3) 
$79^{\circ},0.39$; (4) $-37^{\circ},0.55$; (5) $-117^{\circ},0.72$;
(6) $-82^{\circ},0.88$.}
\label{fig9}
\end{figure}

So far, the angular amplitude $\varphi_{\mathrm{max}}$ of the oscillating
field was always chosen sufficiently small so that the swimmer was moving 
along the $z$ axis. For large $\varphi_{\mathrm{max}}$, however, the
mean velocity $\bar{\bm{v}}$ assumes a non-zero angle $\psi$ with the
$z$ axis. The upper graph of Fig.\ \ref{fig8} summarizes our results.
For $S_{p}$ larger than 7 and angular amplitudes beyond 
$\varphi_{\mathrm{max}} = 70^{\circ}$, the swimming angle $\psi$ jumps 
from $0^{\circ}$ to $90^{\circ}$ and the swimmer thus moves perpendicular to 
the $z$ axis. Such an abrupt transition was also observed in the experiments 
by Dreyfus {\em et al.}\ \cite{Dreyfus05b}. It is associated with a broken
symmetry since the swimmer could also move into the opposite direction
with $\psi = -90^{\circ}$ depending on the initial condition. In the
beginning the filament and magnetic field point along the $z$ axis. When
the magnetic-field direction turns towards the upper-half space ($y>0$),
the filament follows. For large oscillation frequencies $\omega$ 
(large $S_{p}$), it cannot, however, follow the field direction into the 
lower-half space ($y<0$) and $\psi = 90^{\circ}$ results, as pictured
in the third example of Fig.\ \ref{fig9}. For decreasing sperm number, 
this sharp transition becomes smoother involving swimming directions inclined 
relativ to the $z$ axis. Finally for $S_{p} < 3$, the swimmer always swims
along the $z$ axis by following the oscillating field as a nearly rigid 
rod with only small bending.
The simulations of the swimmer's movements for 
$\psi \ne 0$ are very time consuming since the swimmer needs some time 
to reach a steady state. Several snapshots of its configurations for 
different parameters are summarized in Fig.\ \ref{fig9}, their 
locations in the graphs of Fig.\ \ref{fig8} are indicated by filled 
black circles. The temporal evolution of the first and second 
configurations in Fig.\ \ref{fig9} are the dynamic analog of the static 
hairpin structures that form when the aligning magnetic field is suddenly
turned around by $90^{\circ}$\ \cite{Goubault03}. 
Finally, the lower graph of Fig.\ \ref{fig8} plots the swimmming velocity 
$\bar{v}$ as a function of $S_{p}$ and $\varphi_{\mathrm{max}}$. Note that 
the $S_{p}$ and $\varphi_{\mathrm{max}}$ axes are differently oriented 
compared to the upper graph. When the angular amplitude 
$\varphi_{\mathrm{max}}$ increases from $0^{\circ}$, we observe that 
$\bar{v}$ also increases (only partially shown in the graph). At $S_{p} =7$ and 
around $\varphi_{\mathrm{max}} = 70^{\circ}$, a sharp decrease in $\bar{v}$
coincides with the abrupt change of the swimming direction from 
$\psi = 0^{\circ}$ to $90^{\circ}$ in the upper graph. For decreasing
$S_{p}$, the sharp drop in $\bar{v}$ becomes smaller due to the occurence
of inclined swimming directions.

\section{Conclusions} \label{sec.conc}

In this article, we have presented a detailed numerical study of a 
microscopic artificial swimmer realized recently by Dreyfus {\em et al.\/} 
in experiments\ \cite{Dreyfus05}. The elastic superparamagnetic 
filament is modeled by a bead-spring configuration that also resists
bending via a disrete version of the worm-like chain. Friction with the 
surrounding fluid is described by the single beads that also experience
hydrodynamic interactions with each other. The swimmer composed of the 
filament and an attached load is actuated by a magnetic field whose 
direction oscillates.

We show that the superparamagnetic filament, aside from finite-size effects,
can be described by the dimensionless sperm number, the magnitude of the 
magnetic field, and the angular amplitude of the field's oscillating 
direction. We then study the mean velocity and the efficiency of the swimmer,
which we define appropriately, as a function of these parameters and the 
size of the load particle. In particular, we clarify that the real
velocity of the swimmer depends on two main factors namely the shape of 
the beating filament and the oscillation frequency. For given
magnetic-field strength an optimum sperm number (or oscillation frequency)
can be chosen such that mean velocity and efficiency are close to their
maximum values. Whereas the maximum rescaled velocity $\bar{v}/(L\omega)$
as a function of the sperm number only exhibits a weak dependence on the 
magnetic-field strength $B_s$, the real maximum velocity $\bar{v}$ strongly 
increases with $B_s$ since its location is shifted to larger 
$S_{p} \propto \omega^{1/4}$. A study of the influence of the load size
for a particular field strength reveals, the optimum load has to be chosen 
as a compromise between the largest swimming velocity and the best
efficiency. For increasing angular amplitude of the field's oscillating 
direction, the direction of the swimming velocity changes in a 
symmetry-breaking transition that is sharp for large sperm numbers,
becomes smoother for decreasing $S_{p}$, and ultimately vanishes around $S_{p}=3$.
Accordingly, the jump in the swimming angle relative to the symmetry axis
decreases from $90^{\circ}$ to $0^{\circ}$.

\begin{figure}
\includegraphics[width=0.95\columnwidth]{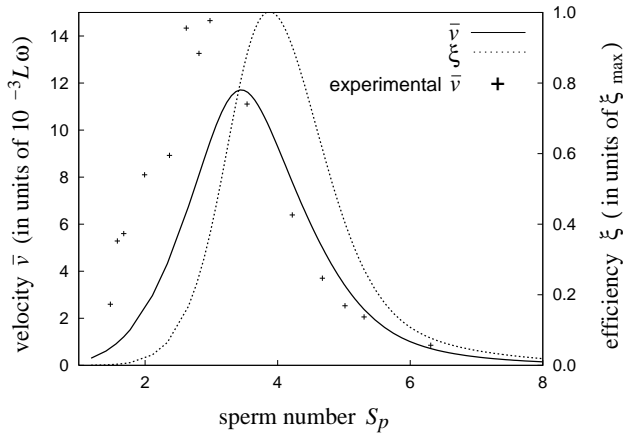}
\caption{Swimming velocity $\bar{v}$ and efficiency $\xi$ in units
of $\xi_{\mathrm{max}} = 1.8 \cdot 10^{-3}$ as a function of 
sperm number $S_{p}$. Parameters are the same as the one used for the
red experimental data points of Fig. 4 in Ref.\ \cite{Dreyfus05}. 
The experimental points are included in this figure with the symbol $+$.}
\label{fig10}
\end{figure}

We also applied the time protocal of Ref.\ \cite{Dreyfus05} (i.e.,
a constant $z$ component and an oscillating $y$ component of the magnetic
field) to actuate the one-armed swimmer. For the parameters of the
red experimental data points of Fig. 4 in Ref.\ \cite{Dreyfus05},
we find nearly quantitative agreement, as illustrated in Fig.\ \ref{fig10}.
To achieve this, we had to rescale the actuating magnetic field 
by a factor of 2.53 to account for the larger distance of the beads in our 
modelling and therefore to compensate for the weaker dipole interaction 
compared to the swimmer in Ref.\ \cite{Dreyfus05}. This makes sense
since the strength of the dipole interaction of neighboring dipoles is the
important parameter for the actuation of the swimmer. Deviations between
our simulations and the experimental results might be due to the fact that
we use a spherical load particle compared to the oblate shape of the
red blood cell used in Ref.\ \cite{Dreyfus05} and that we neglect 
corrections to the actuating external field due to the induced dipole
fields.

In Ref.\ \cite{Japaner}, the authors envisage micro machines moving
in blood vessels and doing necessary repair work. The one-armed swimmer
offers an interesting possibility to propel such micro machines. Our 
numerical studies demonstrate that theoretical modelling helps to 
elucidate the basic features but also to optimize the performance of such 
micro machines.

\begin{acknowledgments}
We would like to thank J. Bibette and R. Dreyfus for making us familiar
with the one-armed swimmer and together with M. Reichert for
helpful discussions. H.S. acknowledges financial support from the 
Deutsche Forschungsgemeinschaft under Grant No. Sta 352/5-2. 
E.G. and H.S. thank the International Graduate College at the University of
Konstanz for financial support.
\end{acknowledgments}

%
%
%
%
%
%

\appendix

\section{Parameter sets} \label{app.par}

In table\ \ref{tab1}, we summarize the parameters in set I and II used for
our numerical studies. The oscillation frequency $\omega = 2\pi/T$
and the magnetic-field strength $B$ were varied to study the respective
ranges of sperm number $S_{p}$ and reduced magnetic field $B_s$. The spring
constant $k$, the time step $\Delta t$ for the Euler integration, and
the number $n_{s}$ of simulation cycles were adjusted as necessary. Typical
values are shown.

\begin{table}
\caption{Parameter sets I and II used in the numerical studies.}
\label{tab1}
\begin{tabular}{c@{\hspace{1.5cm}}c@{\hspace{1.5cm}}c}
\hline
parameter & set I & set II \\
\hline
\hline
$N$ & 20 & 20\\
$a[\mu\mathrm{m}]$ & 0.5 & 2.25 \\
$a_{0}$ & 5$a$ & 8$a$ \\
$l_{0}$ & 3$a$ & 3$a$ \\[1ex] 
$\chi$ & 0.993 & 1.704\\
$\eta [\mathrm{Ns/m^2}]$ & $10^{-3}$ & $10^{-3}$\\[1ex]
$k[\mathrm{N/m}]$ & $1.5\cdot 10^{-3}$  &  1.0 \\
$A[\mathrm{Nm}]$  & $4.5 \cdot 10^{-22}$ &  $6.75 \cdot 10^{-18}$\\[1ex]
$\omega [2\pi/s]\enspace$ & 1 \dots 2285 & 100 \dots $9.4 \cdot 10^{4}$\\
$\enspace\rightarrow S_{p}$ & 2.903 \dots 20.074 & 3.636 \dots 20.134 \\
$B[\mathrm{T}]\enspace$ & $0.07$ & 0.01 \dots 0.49   \\
$\enspace\rightarrow B_{s}$ & 5.76 & 0.233 \dots 11.433 \\
$\varphi_{\mathrm{max}} [\enspace^{\circ}]$ & 45 & 57 \\[1ex]
$\Delta t [\mathrm{s}]$ & $10^{-6}$ & $10^{-8}$ \\
$n_{s}$ & 15 & 15\\
\hline
\end{tabular}
\end{table}




\begin{thebibliography}{99}
\bibitem{twisting}
Since the filament is actuated by an external field that only oscillates
in a plane, twisting of the filament does not occur and, therefore, does 
not have to be taken into account when formulating the free energy of the 
system.
\bibitem{Rotne}
In our modelling, we use a center-to-center distance $3a$ of the particles 
in the filament (where $a$ is the particle radius). In this case, the 
Rotne-Prager approximation deviates from the full multipole results of 
Ref.\ \cite{Cichocki94} by at most 5\%. We checked this by using the program 
Hydrolib (see Ref.\ \cite{Hinsen95}).
\bibitem{footnote}
$B_{s}^{2}$ is proportional to the magnetoelastic number used in Refs.\ 
\cite{Dreyfus05,Roper05}.
\end{thebibliography}
\end{document}